# One-dimensional transport in polymer nanofibers

A. N. Aleshin[1,2,*], H. J. Lee[1], Y. W. Park[1], and K. Akagi[3]

[1]*School of Physics and Nano Systems Institute - National Core Research Center, Seoul National University, Seoul 151-747, Korea*

[2]*A. F. Ioffe Physical-Technical Institute, Russian Academy of Sciences, St. Petersburg 194021, Russia*

[3]*Institute of Materials Science and Tsukuba Research Center for Interdisciplinary Materials Science, University of Tsukuba, Tsukuba, Ibaraki 305-8573, Japan*

We report our transport studies in quasi one-dimensional (1D) conductors - helical polyacetylene fibers doped with iodine and the data analysis for other polymer single fibers and tubes. We found that at 30 K < $T$ < 300 K the conductance and the current-voltage characteristics follow the power law: $G(T) \propto T^{\alpha}$ with $\alpha$ ~ 2.2-7.2 and $I(V) \propto V^{\beta}$ with $\beta$ ~ 2-5.7. Both $G(T)$ and $I(V)$ show the features characteristic of 1D systems such as Luttinger liquid or Wigner crystal. The relationship between our results and theories for tunneling in 1D systems is discussed.

The transport properties of one-dimensional (1D) conductors are a subject of intense interest because of recent development of semiconductor nanotechnology [1]. Electron-electron interactions (EEI) are strongly affecting the transport in 1D systems by leading to phases different from conventional Fermi liquid. The repulsive short-range EEI result in Luttinger liquids (LL) [2], while long-range Coulomb interactions (LRCI) cause a Wigner crystal (WC) [3]. The characteristic feature of 1D systems is the power-law behavior of the tunneling density of states near the Fermi level, which manifests itself in the power-law temperature dependencies of the conductance, *G(T),* and the current-voltage (*I-V*) characteristics, *I(V)*. This behavior has been observed recently in 1D systems, such as single-wall and multi-wall carbon nanotubes (SWNT and MWNT) [4,5], doped semiconductor nanowires [6,7], and in fractional quantum Hall edge states [8]. The LL, WC or Environmental Coulomb blockade (ECB) [5] models are currently debated in order to explain the above-mentioned power-law behavior in 1D systems. Nanofibers that are made of conjugated polymers are another example of initially 1D systems. Polyacetylene (PA) nanofibers are of particular interest because of the unique chemical structure of this polymer. PA has the bandgap of $E_g \sim 1.5$ eV, and consists of linear chains of CH units with alternating single and double bonds. Their conductivity greatly increases upon doping [9]. Both of these features make PA fibers promising for nanotransport studies. The low temperature transport in doped single PA fibers and polypyrrole (PPy) nanotubes has been studied recently [10,11]; the tunneling transport mechanism has also been considered [12]. While the charge transport in doped helical PA has been studied for 3D films [13], the complete understanding of the transport mechanism and the applicability of theories for tunneling in 1 D systems for a single helical PA fiber is still lacking. In this paper we report our transport studies of helical PA fibers doped with iodine. Analysis of early experimental data for a single PA fibers and polymer tubes is included in order to examine the relationships between various theoretical models for 1D conductors.

Helical PA (hel PA) is synthesized using chiral nematic liquid crystal as a solvent of Ziegler-Natta catalyst by K. Akagi et al., following the procedure presented elsewhere [14]. Helical PA is polycrystalline and can be produced with either R- (counterclockwise) or S- (clockwise) type helicity. The typical length of helical PA fibers is of the order of 10 $\mu$m – much longer than that of the traditional PA. A single fiber has a cross-section, typically 40-65 nm (wide) and 130-300 nm (high). R-helical PA fibers are deposited on a Si substrate with a SiO$_2$ layer and gold or platinum electrodes thermally evaporated on the top, 2 $\mu$m apart. The fiber doped with iodine from vapor phase up to the saturation level (1-2 wt %) [15]. The transport measurements are performed in the 2-probe geometry in a vacuum at ~10$^{-5}$ Torr, using displex-osp cryogenic system with a Keithley 6517A electrometer. The conductance of the sample is determined from the Ohmic regime of the *I-V* at each temperature. The conductance measurements for the same sample at 300 K showed the values of the same order of magnitude in both 2-probe and 4-probe geometry.

Fig. 1 shows the AFM image of a R-hel PA fiber, the inset demonstrates the schematic of a two-probe device based on such a R-hel PA nanofiber. The cross-section of this particular fiber is 65 nm × 290 nm, and its length is as long as 10 $\mu$m. Fig. 2 presents the typical *G(T)* dependencies for R-hel PA nanofibers doped with iodine as well as the results of our analysis of early data for iodine doped single PA nanofiber (diameter ~ 20 nm) and PPy nanotube (diameter ~ 15 nm) available from Refs. 10 and 16. As can be seen from Fig.2, the conductance for all types of polymer fibers follows the power-law behavior $G(T) \propto T^\alpha$ starting from room temperature and down to ~ 30 K for the most conductive R-hel PA fiber (sample 1). The power exponent increases as the fiber diameter or its cross-section (the amount of polymer chains) becomes smaller. The inset to Fig. 3 shows that the *I-V* characteristics of a R-hel PA fiber (sample 1) at low temperature also follow the power law $I(V) \propto V^\beta$ with $\beta$ ~ 2.5. The same power-law variations for both *G(T)* and *I(V)* are found for all other polymer fibers and tubes with a variety of power exponents $\alpha$ ~ 2.2 - 7.2 and $\beta$ ~ 2 - 5.7. The variation of $\alpha$ and $\beta$ from one fiber to another may result from the fiber cross section scattering or doping level variation. The parameters for all polymer fibers and tubes under consideration are summarized in Table 1. From these data it emerges that the power-law variations in $G(T)$ and $I(V)$ are a general feature of charge transport in polymer nanofibers and tubes. Note that the similar power-law variations are found for inorganic 1D systems including SWNT and MWNT, ($\alpha$, $\beta \sim 0.36$) [4,5], InSb ($\alpha \sim 2$-$7$, $\beta \sim 2 - 6$) [6] and NbSe$_3$ ($\alpha \sim 1 - 3$, $\beta \sim 1.7 - 2.7$) [7] nanowires, and quantum Hall edge states in GaAs ($\alpha$, $\beta \sim 1.4 - 2.7$) [8]. It is evident that polymer fibers and tubes differ from 1D carbon nanotubes and semiconductor nanowires. Each R-hel PA fiber consists of a number of oriented 1D chains, which are for some extent coupled and disordered as a result of doping. Therefore polymer fibers are in fact quasi-1D and the confinement effects are expected to be not significant. However, the remarkable similarities observed in $G(T)$ and $I(V)$ behavior, in particular with inorganic 1D nanowires, brought us to examine various theoretical models for tunneling in 1D systems.

As a first step we have fitted the $G(T)$ curves to 1D variable-range hopping (VRH) [17] and fluctuation-induced tunneling (FIT) [18] models. The 1D VRH implies $G(T) \propto \exp[-(T_0/T)^p]$ with $p \sim 0.5 - 0.75$ depends on the shape of a power-law density of states near the Fermi level. However, even fit with $p \sim 0.25$ for 3D VRH is too strong to explain the $G(T)$ for polymer fibers. The latter FIT model reveals an unreasonable fitting parameters to get the power law in $G(T)$. Therefore both these models do not fit the data well. It is known that for a 3D disordered systems in the critical regime of the metal-insulator transition the temperature dependence of the resistivity, $\rho(T)$, follows an universal power law: $\rho(T) \approx (e^2 p_F/\hbar^2)(k_B T/E_F)^{-1/\eta} \approx T^{-\gamma}$, where $p_F$ is the Fermi momentum, and $1 < \eta < 3$, i.e. $0.33 < \gamma < 1$ [19]. This model with $\gamma < 1$ can explain the transport in non-oriented 3D R-hel PA films at $T > 30$ K [13]. However, since $\gamma > 1$ in the case of polymer fibers, this model is ruled out. The power law $I(V) \propto V^\beta$, with $\beta \approx 2$ is known for the space-charge limited current (SCLC) transport in semiconductors [20]. The values $\beta > 2$ observed for R-hel PA fibers and PPy tubes are inapplicable to the SCLC model. At last, as mentioned above, the power-law variations of $G(T)$ and $I(V)$ in inorganic 1D systems, such as SWNT, MWNT, doped semiconductor nanowires are discussed in terms of LL, WC or ECB theories [4-8]. For a clean LL a power-law variation $G(T) \propto T^\alpha$ is predicted at small biases ($eV \ll k_B T$), and $I(V) \propto V^\beta$ - at large biases ($eV \gg k_B T$). The exponents of the power law depend on the number of 1D channels [21]; the LL state survives for a few 1D chains coupled by Coulomb interactions and can be stabilized in the presence of impurities for more than two coupled 1D chains [22]. According to the LL theory, $I$-$V$ curves taken at different temperatures should be fitted by the general equation [4,23]:

$$I = I_0 \, T^{\alpha+1} \sinh(eV/k_B T) \, |\Gamma(1 + \beta/2 + i\, eV/\pi k_B T)|^2 \qquad (1)$$

where $\Gamma$ is the Gamma function, $I_0$ is a constant. The parameters $\alpha$ and $\beta$ correspond to the exponents estimated from $G(T)$ and $I(V)$ plots. It means that $I$-$V$ curves should collapse into a single curve if $I/T^{\alpha+1}$ is plotted as a function of $eV/k_B T$. As can be seen from Fig. 3, all thus scaled $I$-$V$ curves for a R-hel PA fiber (sample 1) collapse into a single curve at low temperatures by plotting $I/T^{\alpha+1}$ versus $eV/k_B T$, where $\alpha \sim 2.8$ is the power-law exponent from the $G(T)$ curve for the same sample. The similar scaled behavior is found for all other R-hel PA fibers as well as for a single PA fiber and PPy tubes. We conclude that the power-law variations in $G(T)$ and $I(V)$ and the scaled $I$-$V$ behavior are characteristic of such quasi-1D systems as polymer fibers and tubes. To a certain degree this behavior correlates with a LL theory for transport in 1D systems, which implies that both: tunneling along LLs through impurity barriers and tunneling between the chains with LLs provides the conduction of a set of coupled LLs. The LL model predicts the correlation between the exponents $\beta = \alpha + 1$, however for all polymer fibers $\beta \neq \alpha + 1$, and, moreover, $\beta$ is always less than $\alpha$, which can not be explained either LL or ECB theory [5]. Note that the same disagreement with the LL model is found earlier for inorganic 1D nanowires [6,7]. The LL model implies almost purely 1D transport, while the interchain hopping destroys the LL state. In view of the interchain hopping to be one of the main transport mechanism in conducting polymers at low temperature [9], one can expect the manifestation of the LL state in polymer fibers at relatively high temperatures only. This correlates with our results obtained for R-hel PA nanofibers at $T > 30$ K.

On the other hand, the consideration of repulsive LRCI is believed to cause a WC, which is pinned by impurity [3,24]. WC occurs in solids with a low electronic density, i.e. with a large parameter $r_S = a/2a_B$, where $a$ is the average distance between electrons and $a_B$ is the effective Bohr radius. The WC transition has been shown to take place at $r_S \sim 36$ [25]. Such a pinned WC can adjust its phase in the presence of disorder to optimize the pinning energy gain, similar to classical charge density wave systems. Polymer nanofibers and tubes are quasi-1D conductors with a low electronic density, where WC may occur. Because the molecular structure of PA consists of alternating single and double bounds in the CH chain those results in the Peierls distortion. This creates the potential wells where the electrons can get trapped in and hence form the WC (see the inset to Fig. 2). At the modest doping level of the PA chain the distance between impurities (iodine) is of the order 10 nm, so $r_S$ is large enough to get WC in the PA fiber. When the localization length is larger than the distance between impurities the tunneling density of states in WC follows a power law [3,24,26] with the exponents $\sim 3 - 6$ [27], similar to those found in $G(T)$ for polymer nanofibers. However, the impurities in doped conjugated polymers are sited outside of the polymer chains and therefore they only supply charge carriers. This result in the polymer chain either not pinned or pinned weakly by conjugated defects. For a classical 1D WC pinned by impurities one should expect the VRH regime at low temperature with an exponential $G(T)$ [17,28]. The absence of effective pinning prevents the observation of VRH in polymer fibers at least at $T > 30$ K and argues against the WC model; however this regime may manifest itself at lower temperatures.

Therefore, all models under consideration, namely VRH, LL, ECB and WC, can not describe precisely the power-law variations observed in $G(T)$ and $I(V)$ for polymer nanofibers despite of some relationship with LL and WC models is found. We suppose that the real transport mechanism in quasi-1D polymer fibers obeys ether the superposition of the above-mentioned models or the single LL-like model valid for the different parts of the metallic polymer fiber separated by intramolecular junctions. By analogy with the LL transport in bent metallic carbon nanotube [29], we suggest that the conductance across the intramolecular junction is much more temperature dependent than that of the two (or more) straight segments, but still obeys the power law $G(T) \propto T^\alpha$. In the case of SWNT for the end-to-end tunneling between two LLs separated with intramolecular junction (kink) the power exponent $\alpha_{\text{end-end}} = (1/g - 1)/2 \sim 1.8$, where the LL interaction parameter $g \sim 0.22$. For the SWNT with a single kink $\alpha = 2.2$ [29], which is close to the calculated value and surprisingly close to the $\alpha$ value for the most conductive R-hel PA fiber (sample 1). It is evident from the AFM image (Fig.1), that R-hel PA fiber contains kinks at a distance between electrodes 2 μm. In our opinion each kink in the polymer fiber acts as the tunneling junction between the ends of two LLs. This results in stronger power-law variations in $G(T)$ and $I(V)$ with respect to those in a clean LL, and may caused the above-mentioned contradictions with a pure LL model. In conclusion, we have found that at 30 K $< T < 300$ K the conductance and the current-voltage characteristics of iodine doped helical

PA fibers and other conjugated polymer fibers and tubes follow the power laws characteristic of 1D systems, such as LL or WC. At the same time there is a discrepancy between our results and theories for tunneling in 1D systems.

The authors are grateful to B. I. Shklovskii and S. Brazovskii for comments. This work was supported by KISTEP, the contract M6-0301-00-0005, Korea. Support from the Brain Pool Program of KFSTS for A. N. A. is gratefully acknowledged.

**Table 1. Polymer fibers and tubes used in transport experiments**

| # | Sample | Diameter or cross-section nm | $G_{300K}$ S | $\alpha$ | $\beta$ (at min T) | Reference |
|---|---|---|---|---|---|---|
| 1 | R-hel PA fiber | 65×290 | $8.4\times10^{-7}$ | 2.2<br>2.8 | -<br>2.5 (30 K) | this work<br>- " - |
| 2 | R-hel PA fiber | 60×134 | $1.1\times10^{-7}$ | 5.5 | 4.8 (50 K) | - " - |
| 3 | R-hel PA fiber | 47×312 | $2.1\times10^{-9}$ | 7.2 | 5.7 (95 K) | - " - |
| 4 | Single PA fiber | 20 | $7.3\times10^{-9}$ | 5.6 | 2.0 (94 K) | [10] |
| 5 | PPy tube | 15 | $1.7\times10^{-8}$ | 5.0 | 2.1 (56 K) | [16] |
| 6 | PPy tube | 50 | $2.8\times10^{-8}$ | 4.1 | 2.8 (50 K) | [16] |

**Figures captions.**

Fig. 1 AFM image of a R-hel PA fiber, the inset shows the schematic of a two-probe device based on such a R-hel PA fiber on top of Pt electrodes.

Fig. 2 Conductance vs $T$ for iodine doped R-hel PA fibers, iodine doped single PA fiber (diameter ~ 20 nm) and PPy nanotube (diameter ~ 15 nm) (samples 1-5 from Table 1). Inset shows the coupled double-well potentials with a conjugated defect in PA chain.

Fig. 3 $I/T^{\alpha+1}$ vs $eV/k_BT$ for R-hel PA fiber (sample 1) at different temperatures, $\alpha$ ~2.8 is the exponent in $G(T) \propto T^{\alpha}$. The inset shows $I$-$V$ curve for sample 1 at $T = 30$ K.

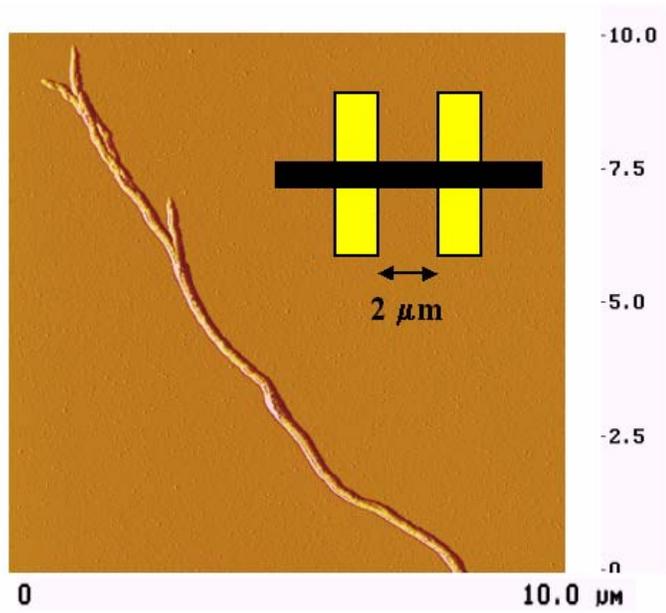

Fig. 1  A. N. Aleshin *et al.*

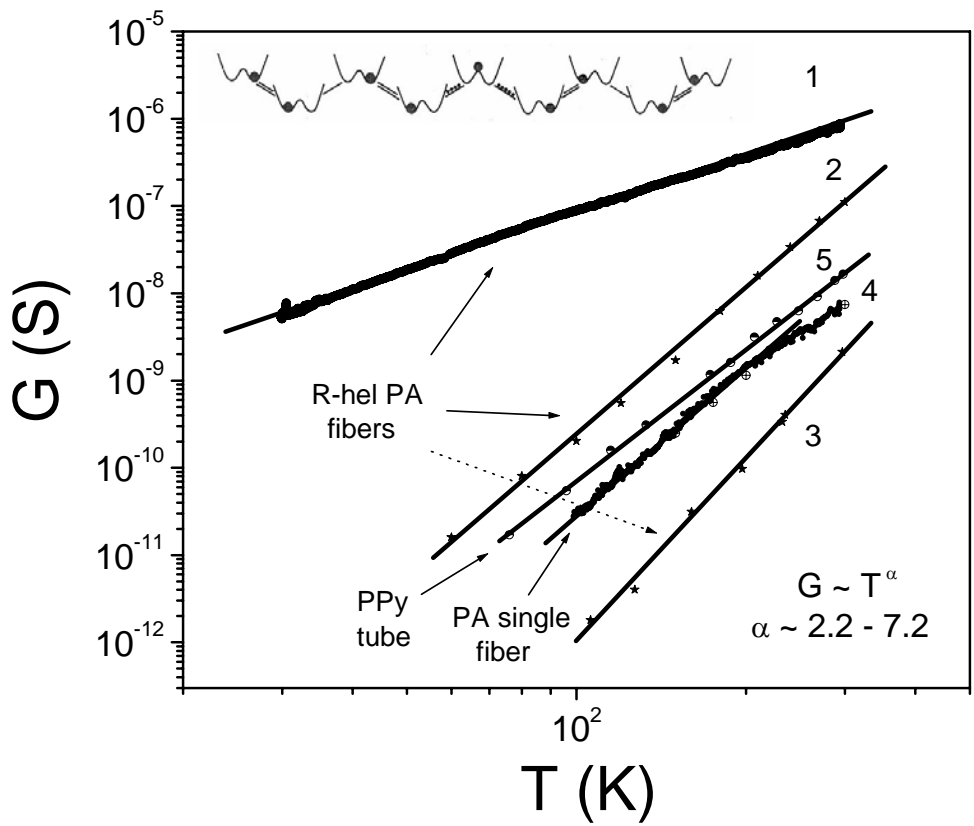

Fig. 2 A. N. Aleshin *et al*.

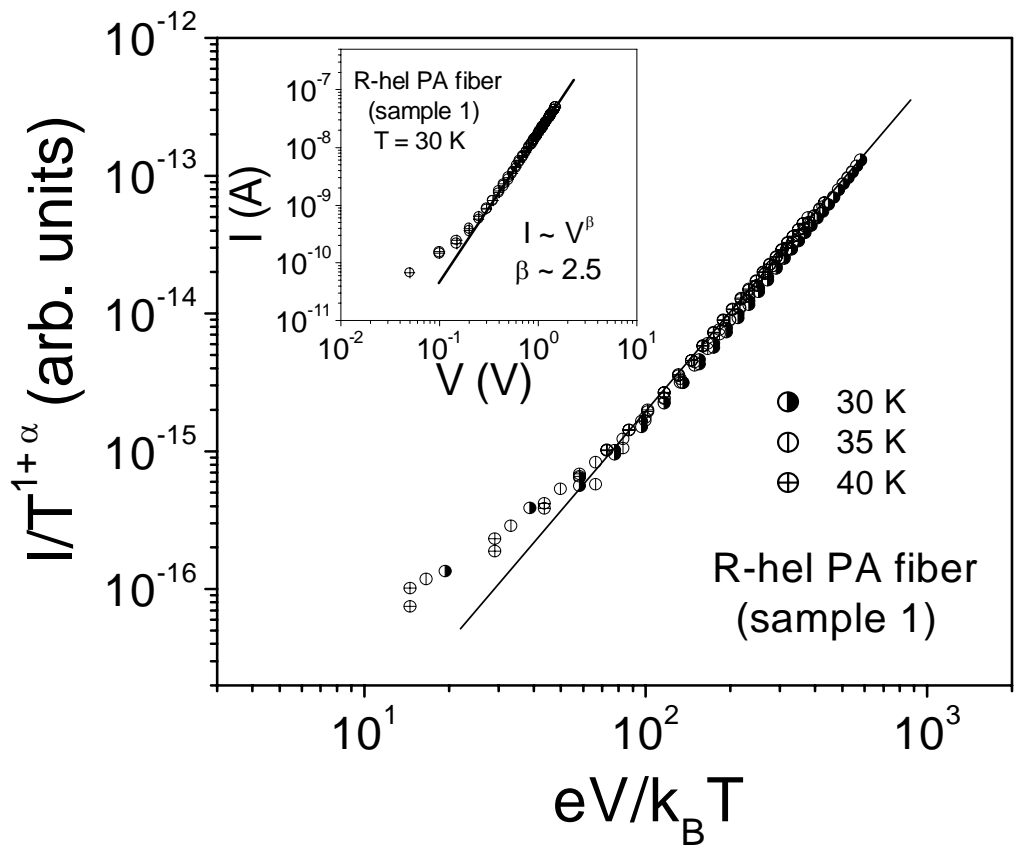

Fig. 3  A. N. Aleshin *et al.*